\def\beg{\begin{equation}}
\def\eeq{\end{equation}}
\documentstyle[12pt]{article}
\begin{document}
\begin{center}
{\Large{\bf Nonuniqueness of Kivelson, Kallin, Arovas and Schrieffer's fractional charge.}}
\vskip0.35cm
{\bf Keshav N. Shrivastava}
\vskip0.25cm
{\it School of Physics, University of Hyderabad,\\
Hyderabad  500046, India}
\end{center}

It is found that the magnetic length has not been treated correctly to calculate the classical action. In fact, the charge and the magnetic length have not been resolved. It is of serious consequences, because fractional charge completely disappears and only the flux area, $\l_o^2$ becomes fractional. The results of Kivelson et al are therefore not unique.
\vfill
Corresponding author: keshav@mailaps.org\\
Fax: +91-40-3010145.Phone: 3010811.
\newpage
\baselineskip22pt
\noindent {\bf 1.~ Introduction}

     Kivelson et al$^1$ suggest that for certain filling factors, $\nu_c=n/m$, the ratio of two integers, the energy is stable so 
that the Wigner crystal is unstable for $\nu \neq \nu_c$. We find 
that this result is obtained by ignoring $\l_o$ which is the 
magnetic length. When this magnetic length is considered properly, 
the need for the fractionally charged particles disappears. But 
there are fractionally charged quasiparticles, then what they are 
due to?

     In this paper, we correct the paper of Kivelson et al. We 
introduce the magnetic length where it was left out but then the interpretation changes. Therefore, we are also able to correct the interpretation.

\noindent{\bf 2.~~Theory}

      The path-integral representation of the partition function, 
$Z= Tr \,\,exp (-\beta H_N)$ is obtained from the $N$ particle hamiltonian,
\beg
{\it H}_N=\sum_{i=1}^N{1\over{2m^*}}[p_i+{e\over {2c}}{\vec B}_o
{\hat z}\times \vec r_i]^2+\sum_{j<k}V_2(\vec r_j - \vec r_k).
\eeq
The lowest Landau level (LLL) is given by,
\beg
\phi_R(r) = (2\pi)^{-1/2}exp[-{1\over 4}(\vec r - \vec R)^2+
{1\over 2} i(\vec r \times \vec R)\times \hat z ].
\eeq

Since Kivelson et al suggest $\l_o=1$, the above wave function
is dimensionally not correct, so it may be assumed that when 
necessary the dimensions will be corrected. The path integral 
representation for $Z$ is given by,
\beg
Z(\nu)=(1/N!)\Sigma_{P\epsilon S_N}(-1)^P\int d^{2N}r 
<j|exp[-\beta {\it H}_N]|P(i)>
\eeq
where,
\beg
 N= \nu B_o/\phi_o.
\eeq
The inconsistencies in the formulas as well as in discussions 
are clearly visible. In (3), $N$ is a pure number so that its 
factorial is well defined but in (4) $N$ is actually a number per 
unit area, i.e., the number density but the factorial is not defined 
for the number density which need not be an integer. The action for 
a continuous path is defined as $\int {\it L} dt$ where ${\it L}$
is the Lagrangian of the system as,
\beg
S[R] =(1/2)\int_o^{\beta} d\tau[-i\Sigma_{j=1}^N(\dot{\bf R_j}\times {\bf R}_j).\hat z+ \Sigma_{j\neq k}V({\bf R}_j-{\bf R}_k)]
\eeq
where $V$ is the matrix element of the Coulomb potential between coherent states with $\tau$ as the imaginary time,
\beg
V(R) =(1/2){\sqrt \pi}(e^2/\epsilon)exp(-R^2/8)I_o(R^2/8).
\eeq
Here the prefactor on the right hand side does not have the 
dimensions of a potential energy. The factor $e^2/\epsilon$ should 
be replaced by $e^2/(\epsilon \l_o)$. This correction is very 
important because the correction can now occur in $\l_o$ otherwise 
only the charge can be corrected. Similarly, the argument of the exponential function requires to be corrected. The partition function $Z$ is evaluated by finding all paths $R^c(\tau)$ for which the 
action has an extremum value. Here $R^c(\tau)$ is a vector function 
with 2$N$ components $R_j(\tau)$. Again, the number of components is 
not 2$N$ because $N$ should be treated as density. The number of electrons in the ring is $L$. The real part of the action is $\alpha_o(\nu)L$ and the imaginary part determines the time dependence.
Kivelson et al suggest that the relaxation corrections are small but there is no particular reason to think that such quantities should be compared at all. For a phase transition $\alpha(\nu)<\alpha_c$ some
critical value. It has been suggested that this inequality is most 
likely satisfied by $\nu$= 1/3, 1/5, ..., 4/9, but the fractions 
1/3 etc are not derived.

Suppose that there is a local dilation of the Wigner crystal by an amount $\delta A$, then the phase acquired is $\Delta \theta = 2\pi B_o\delta A/\phi_o$. Kivelson et al suggest that this means that the quasiparticle charge is,
\beg
Q^* = \pm \nu e,
\eeq
because one can absorb $\pm \nu$ in $\phi_o$=hc/e. Here, 
$\delta Ae/hc$ is the quantity which is occuring in the algebra so 
that $\pm \nu$ need not be absorbed in $e$ to change it to 
$\pm \nu e$. Instead of absorbing $\pm \nu$ in $e$, we can absorb 
it in $\delta A$. The charge $e$ then remains unchanged and 
$\delta A$ changes to $\pm \nu \delta A$. Kivelson et al's result, $Q^*=\pm\nu e$, is then not necessarily correct.

Kivelson et al have estimated the energy for creation of a 
quasiparticle as,
\beg
E_{qp}(\nu)\sim 0.5\nu^2e^2/\epsilon_o.
\eeq
This is not having the correct dimensions. To set it right, we can change it to $E_{qp}(\nu)\simeq 0.5\nu^2e^2/\epsilon_o\l_o.$ Then 
charge can be $e$ and only $\l_o$ is changed to $\l_o/\nu^2$. 
Therefore, the arguments used to discuss the creation energy are 
not satisfactory. Similarly, it is found that the arguments used to discuss the fractional charge in Peierls distortion and in the calculation of Berry's phase are not correct.
 
\noindent{\bf 3.~~Conclusions}

     Kivelson et al have calculated the classical action from 
which they claim that the quasiparticles are fractionally charged. 
We have checked their calculation and find that the quasiparticles 
need not be fractionally charged and Kivelson's results are not 
unique. There may be a fractional area instead of the fractional 
charge.

Schrieffer et al have published several papers dealing with the fractional charge. It has been reported that in the case of a Peierls distortion, the charge may be fractional. A calculation of the Berry's phase has been published and in the prsent case, the classical action has been calculated. In all the cases, a fractional charge has been reported. Upon closer scrutiny of the algebraic derivations, it has been found that the calculation has not been performed correctly. We have found that the calculations have errors. Su and Schrieffer$^2$ have discussed the Peierls distortion and Arovas et al$^3$ have calculated the Berry's phase. We have shown$^4$ that both of these papers of Schrieffer are in error. Similarly, Laughlin's paper$^5$ is also not correctly written. The error has been made in such a way that ``exactness" is not affected$^6$.                                                                             
                                                                                  
The correct theory of the quantum Hall effect is given in ref.7

{\it About the author: Keshav Shrivastava obtained his B.Sc. from Agra University, M.Sc. from the Allahabad University, Ph.D. from the Indian Institute of Technology, D.Sc. from the Calcutta University and F. Inst. P. from U.K. He has published 170 papers and two books. He has discovered the flux quantized energies in superconductors and correct fractional charges in the quantum Hall effect.}

\noindent{\bf4.~~References}
\begin{enumerate}
\item S. Kivelson, C. Kallin, D. P. Arovas and J. R. Schrieffer, Phys. Rev. Lett. {\bf 56}, 873 (1986).
\item W. P. Su and J. R. Schrieffer, Phys. Rev. Lett. {\bf46}, 738 (1981).
\item D. Arovas, J. R. Schrieffer and F. Wilczek, Phys. Rev. Lett. {\bf53}, 722 (1984).
\item K. N. Shrivastava, cond-nmat/0211351
\item R. B. Laughlin, Phys. Rev. Lett. {\bf50}, 1395(1983).
\item K. N. Shrivastava, cond-mat/0210238.
\item K. N. Shrivastava, Introduction to quantum Hall effect,\\ 
      Nova Science Pub. Inc., N. Y. (2002).
\end{enumerate}
\vskip0.1cm
Note: Ref.7 is available from:\\
 Nova Science Publishers, Inc.,\\
400 Oser Avenue, Suite 1600,\\
 Hauppauge, N. Y.. 11788-3619,\\
Tel.(631)-231-7269, Fax: (631)-231-8175,\\
 ISBN 1-59033-419-1 US$\$69$.\\
E-mail: novascience@Earthlink.net

\vskip0.5cm

\end{document}